\documentclass{optica-article}
\usepackage{enumitem} 
\usepackage[separate-uncertainty=true]{siunitx}

\journal{opticajournal} 

\articletype{Research Article}

\usepackage{lineno}

\begin{document}

 \title{Widely non-degenerate nonlinear frequency conversion in cryogenic titanium in-diffused lithium niobate waveguides}

\author{Nina Amelie Lange,\authormark{1,2,*} Sebastian Lengeling,\authormark{3} Philipp Mues,\authormark{2} Viktor Quiring,\authormark{3} Werner Ridder,\authormark{3} Christof Eigner,\authormark{2} Harald Herrmann,\authormark{3} Christine Silberhorn,\authormark{3} and Tim J. Bartley\authormark{1,2}}

\address{\authormark{1}Department of Physics, Paderborn University, Warburger Str. 100, 33098 Paderborn, Germany\\
\authormark{2}Institute for Photonic Quantum Systems (PhoQS), Paderborn University, Warburger Str. 100, 33098 Paderborn, Germany\\
\authormark{3}Integrated Quantum Optics Group, Institute for Photonic Quantum Systems (PhoQS), Paderborn University, Warburger Str. 100, 33098 Paderborn, Germany}

\email{\authormark{*}nina.amelie.lange@upb.de} 


\begin{abstract*} 
The titanium in-diffused lithium niobate waveguide platform is well-established for reliable prototyping and packaging of many quantum photonic components at room temperature. Nevertheless, compatibility with certain quantum light sources and superconducting detectors requires operation under cryogenic conditions. We characterize alterations in phase-matching and mode guiding of a non-degenerate spontaneous parametric down-conversion process emitting around \SI{1556}{nm} and \SI{950}{nm}, under cryogenic conditions. Despite the effects of pyroelectricity and photorefraction, the spectral properties match our theoretical model. Nevertheless, these effects cause small but significant variations within and between cooling cycles. These measurements provide a first benchmark against which other nonlinear photonic integration platforms, such as thin-film lithium niobate, can be compared.
\end{abstract*}

\section{Introduction}
Lithium niobate is a well-established platform for integrated quantum circuits, due to its large nonlinear and electro-optic coefficients, a broadband transparency window, and ferroelectric properties~\cite{weis1985lithium,sohler2008integrated,alibart2016quantum,chen2022advances}. Its strong second-order nonlinearity allows for efficient frequency conversion processes~\cite{regener1988efficient,fujii2007bright,boyd2008nonlinear,wang2018ultrahigh,stefszky2018high}. Compared to bulk crystals, waveguides in lithium niobate enable confinement of the interacting electromagnetic fields and thus stronger mode overlap, which can result in significantly higher brightness~\cite{sohler2008integrated}. Since lithium niobate is ferroelectric, periodic poling is a common method to account for the natural phase-mismatch of the individual modes (i.e. quasi-phase-matching)~\cite{yamada1993first,tanzilli2002ppln,niu2022research}. Careful selection and accurate fabrication of the poling period enable high conversion efficiency of a desired wavelength interaction.

The fabrication of periodically poled titanium in-diffused waveguides in $z$-cut lithium niobate (Ti:PPLN) is an established technology for realizing various integrated devices~\cite{alferness1988titanium}, such as photon-pair sources~\cite{martin2010polarization,krapick2013efficient,montaut2017high,sansoni2017two,pollmann2024integrated}, passive and active manipulation of quantum states~\cite{alferness1979characteristics,huang2007electro,luo2019nonlinear,thiele2022cryogenic}, and single-photon detectors~\cite{hopker2019integrated,hopker2021integrated}. Nowadays, lithium niobate on insulator (LNOI), also known as thin-film lithium niobate (TFLN), represents a very promising platform for integrated optics~\cite{zhu2021integrated}. Strongly confined waveguide modes allow for compact circuit designs with small bending radii and high nonlinear conversion efficiencies. In comparison to LNOI, in-diffused waveguides in lithium niobate exhibit weaker mode confinement, which directly limits the nonlinear conversion efficiency~\cite{chen2022advances}. However, this weak confinement is notably beneficial for free-space waveguide coupling and direct fiber-to-chip packaging, making Ti:PPLN well-suited for integrated devices that do not rely on a high density of components. Especially in the telecom range, the waveguide mode exhibits a large overlap with the mode size of commercially available single-mode optical fibers~\cite{alferness1982efficient,montaut2017high,hopker2019integrated}. This allows for high coupling efficiencies without the need of additional mode matching, e.g., by the use of tapers. While the fabrication of grating couplers with large spectral bandwidth represents an ongoing challenge for TFLN~\cite{zhou2023high,he2024inverse,zhou2025waveguide}, direct free-space coupling to Ti:PPLN enables broadband waveguide coupling, which is key for realizing widely non-degenerate frequency conversion processes. The well-researched and optimized manufacturing process also facilitates prototyping, whilst free-space waveguide coupling allows for rapid, sequential investigation of multiple waveguides on one chip.

Interfacing a large number of optical components under mutually compatible operating conditions is essential for scalable quantum photonics. While a multitude of integrated components is optimized for room temperature operation in lithium niobate, certain devices, such as highly-efficient superconducting detectors, or quantum light sources based on solid state quantum emitters, require cryogenic operation~\cite{eisaman2011invited,elshaari2020hybrid}. Interfacing these devices demands that all combined components are compatible, which means in this case, cryogenically compatible. For this reason, it is crucial to expand the toolbox of cryogenic integrated photonics. Research in recent years has shown an increasing demand for the development of cryogenic components which are integrated on different material platforms. Among others, Ti:PPLN has been demonstrated to enable cryogenic integration of nonlinear frequency conversion, including second harmonic generation (SHG)~\cite{bartnick2021cryogenic}, and spontaneous parametric down-conversion (SPDC) in the telecom C-band~\cite{Lange2022cryogenic, lange2023degenerate}, electro-optic modulators (EOMs) for phase modulation, directional coupling, and polarization conversion~\cite{thiele2020cryogenic, thiele2022cryogenic}, and evanescently coupled superconducting single-photon detectors~\cite{hopker2019integrated,hopker2021integrated}. TFLN has shown great performance for cryogenic operation, including nonlinear processes such as SHG and SPDC in the telecom C-band~\cite{cheng2024efficient}, a spectral optical filter~\cite{Cheng2024cryogenic}, and even combined integration of EOMs and superconducting nanowire single-photon detectors (SNSPDs) on a single LNOI chip~\cite{lomonte2021single}. Moreover, it has been demonstrated that silicon waveguides enable cryogenic single-photon generation via spontaneous four-wave mixing~\cite{Feng2023entanglement, witt2024packaged}, and graphene allows for efficient EOM performance due to an intrinsic carrier mobility increase at low temperatures~\cite{lee2021high}. 

We have shown before the proof-of-principle functionality of cryogenic integrated type-II SPDC in our waveguide platform~\cite{Lange2022cryogenic}, and that we can generate degenerate photon pairs in the telecom C-band from our cryogenic source via precise fabrication of the required poling period~\cite{lange2023degenerate}. However, we have also seen, that our sources exhibit reduced efficiencies under cryogenic operation. We mainly attribute these observations to the influence of the photorefractive and pyroelectric effects. Both effects can cause an inhomogeneous accumulation of electric charges resulting in refractive index changes, thus disturbing the waveguiding itself, or the phase-matching condition~\cite{rams2000optical,parravicini2011all,pal2015photorefractive,bartnick2021cryogenic,thiele2024pyroelectric}. Understanding these aspects is crucial for interfacing Ti:PPLN with elements such as superconducting detectors to achieve on-chip quantum light generation and single-photon detection.

Our previous results were obtained from type-II SPDC sources pumped with transverse electric (TE) polarized light around \SI{780}{nm}, generating orthogonally polarized photons. Since the local electric fields build up along the crystal axis, they mainly affect the transverse magnetic (TM) polarized light, which is polarized along the crystalline $c$-axis in our $z$-cut samples~\cite{chen1969optically,chon1993photorefractive,pal2015photorefractive}. Moreover, the photorefractive effect becomes considerably more prominent when pumped with high-energy photons~\cite{chen1969optically,pal2015photorefractive}. To counteract photorefraction, lithium niobate crystals are often actively heated to increase the charge carrier mobility, raising the threshold of how much optical energy can be coupled in without inducing photorefractive damage~\cite{rams2000optical}. Under cryogenic conditions, this approach is not an option and alternatives must be sought.

To deepen the understanding of how cryogenic operation affects the material and to learn about possible performance boundaries, in this paper we actively pick a nonlinear interaction that is more strongly affected by photorefraction and pyroelectricity than our previous research. We investigate a phase-matched type-0 interaction that employs only TM-polarized light and that is pumped in the visible regime around \SI{590}{nm}. 
The pump photons thus have higher energy than in our previously characterized type-II process, making this type-0 process more sensitive to photorefractive damage. 
This means we rely on successful mode guiding for TM polarization and on keeping the light intensity of the high-energy beam at a low level. 
The waveguide crystal under test is positioned in a free-space coupled cryostat. We characterize the cryogenic phase-matching properties in two steps: first via sum frequency generation (SFG), and second through SPDC to generate photon pairs around \SI{1556}{nm} and \SI{950}{nm}. This wavelength combination can be of interest for quantum teleportation experiments to transfer the information of a photon from a quantum dot to the telecom range~\cite{bouwmeester1997experimental,polyakov2011coalescence,huber2017interfacing}. This study provides insights into the cryogenic properties of lithium niobate and these results may be helpful in the future development of cryogenic integrated devices.

This paper is organized as follows. First, we provide an overview about frequency conversion in lithium niobate waveguides. Thereby, we introduce our waveguide design, and point out the main challenges which must be considered for cryogenic operation. We then show our experimental methods and describe the pump light generation, which is performed by SFG in a separate, actively heated waveguide. The investigation of the cryogenic waveguide is started by characterizing the linear waveguiding performance and the nonlinear phase-matching spectrum. Finally, we explore the pump power dependence and spectral properties of the cryogenic SPDC source.

\section{Frequency conversion in lithium niobate waveguides}
The interacting fields for nonlinear frequency conversion processes must fulfill energy and momentum conservation~\cite{couteau2018spontaneous}. Considering parametric down-conversion, the total energy of the photon pair is defined by the energy of the pump beam, according to $\omega_\mathrm{p} = \omega_\mathrm{s} + \omega_\mathrm{i}$, where $\omega$ is the photon frequency and the subscripts p, s, i denote the pump, signal, and idler field. The phase-matching function reflects momentum conservation and determines how this energy is distributed between the signal and the idler photons. This is specified by the crystal dispersion through the propagation constants $k(\lambda)=2\pi n(\lambda)/\lambda$, with $n$ the effective refractive index of the waveguide mode at the wavelength $\lambda$. For efficient interaction, the phase-mismatch $\Delta k$ in the propagation constants must disappear. Since we employ periodic poling for quasi-phase-matching, the phase-mismatch is moreover modified by the additional term $2\pi/\Lambda$, which allows tunability with the fabricated poling period $\Lambda$. For cryogenic operation, we must consider the temperature dependence of the refractive indices $n(\lambda) \xrightarrow{} n(\lambda,T)$ and the crystal length, i.e., the poling period $\Lambda \xrightarrow{} \Lambda(T)$. The phase-mismatch at a specific crystal temperature $T$ can thus be described by
\begin{equation}
    \Delta k(T) = k_\mathrm{p}(\lambda_\mathrm{p},T) - k_\mathrm{s}(\lambda_\mathrm{s},T) - k_\mathrm{i}(\lambda_\mathrm{i},T) \pm \frac{2\pi}{\Lambda(T)} \,.
    \label{eq:phasemismatch1}
\end{equation}

We set our poling period to phase-match a type-0 nonlinear process for which all three interacting fields are TM polarized. This interaction exploits the strong $d_{33}$ nonlinear coefficient, which results in highest conversion efficiency compared to other processes~\cite{sohler2008integrated,sharapova2017toolbox}. On the other hand, the TM mode travels more closer to the waveguide surface than the TE mode~\cite{hopker2019integrated}. This mode is thus known to experience higher optical losses due to tiny scratches or particles on the surface. Moreover, this polarization is more sensitive to refractive index changes induced by charge carriers since it is oriented in the same direction as arising electric fields and, thus, coupling via the strong electro-optic coefficient $r_{33}$. Through the investigation of a type-0 TM-polarized process, we thus obtain significant insights into the role of electric charge build-up under cryogenic operation and its effect on frequency conversion.

\subsection{Waveguide design}
\begin{figure} [t]
    \centering
    \includegraphics[width=1\linewidth]{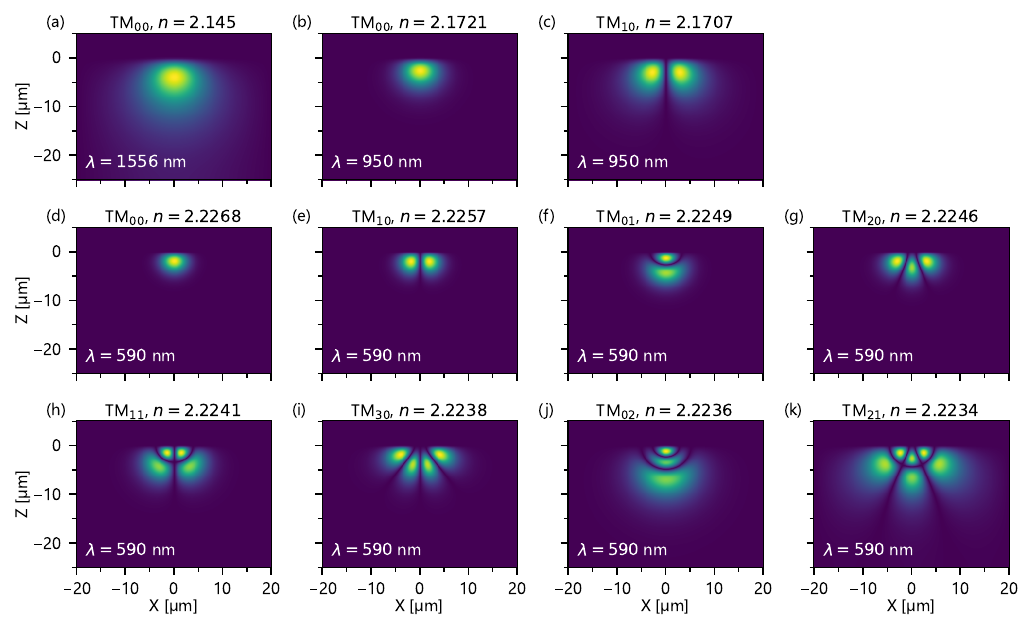}
    \caption{Simulated spatial modes which are guided in \SI{5}{\um} wide Ti:PPLN waveguides at an operation temperature of \SI{175}{\degreeCelsius}. (a) At \SI{1556}{nm} the waveguide is single mode and only allows the fundamental mode, (b) - (c) in addition to the fundamental mode, one higher order mode is guided at \SI{950}{nm}, and (d) - (k) the waveguide becomes highly multi-mode at \SI{590}{nm}.}
    \label{fig:higherModes_RSoft}
\end{figure}

We investigate Ti:PPLN waveguides that are fabricated in house. The titanium stripes are patterned on a $z$-cut lithium niobate substrate with mask lithography and afterwards in-diffused in an oven. The periodic poling is performed with laser lithography. We choose a small waveguide size, defined by Ti-stripes of \SI{5}{\um} width and \SI{80}{nm} thickness (prior to the in-diffusion), to limit the number of higher order spatial modes. We simulate the spatial mode profile of the guided modes using a commercial mode solving software based on the finite-element method (RSoft FemSIM). This simulation takes into account, among others, the refractive index of bulk lithium niobate~\cite{edwards1984temperature,jundt1997temperature}, together with the dimensions of the titanium stripe and its diffusion parameters. While the waveguides are single mode in the telecom wavelength range, they become multi mode for shorter wavelengths. We show the simulated spatial modes for the three interacting wavelengths \SI{1556}{nm}, \SI{950}{nm}, and \SI{590}{nm} in Fig.~\ref{fig:higherModes_RSoft}. Since the chosen wavelengths are widely non-degenerate and we consider waveguides with low confinement, there is a strong difference in size for their fundamental modes $\mathrm{TM}_{00}$ (see Fig.~\ref{fig:higherModes_RSoft}~(a), (b), (d)). The overlap of these modes is \SI{23.0}{\%}, which limits the overall achievable conversion efficiency. At a wavelength of \SI{950}{nm} the waveguides support only one higher order spatial mode $\mathrm{TM}_{10}$ in addition to the fundamental one (see Fig.~\ref{fig:higherModes_RSoft}~(c)), while the smaller mode size in the visible range results in guiding of seven higher order modes (see Fig.~\ref{fig:higherModes_RSoft}~(e)~-~(k)).

Each spatial mode experiences a slightly different effective refractive index which results in different phase-matching conditions. As the fundamental modes of the interacting fields show the largest spatial overlap, we concentrate on phase-matching for this process. This is indicated by using the notation $n_\mathrm{TM00}$ when modifying Eq.~(\ref{eq:phasemismatch1}) to find the required poling period to achieve phase-matching, i.e., $\Delta k = 0$,

\begin{equation}
     \frac{1}{\Lambda(T)} = \frac{n_\mathrm{TM00}(\lambda_\mathrm{p},T)}{\lambda_\mathrm{p}} - \frac{n_\mathrm{TM00}(\lambda_\mathrm{s},T)}{\lambda_\mathrm{s}} - \frac{n_\mathrm{TM00}(\lambda_\mathrm{i},T)}{\lambda_\mathrm{i}} \,.
     \label{eq:phasemismatch2}
\end{equation}
The temperature dependence of this equation is often exploited to tune the phase-matching point of a Ti:PPLN waveguide by changing the crystal temperature. This allows one to compensate for fabrication imperfections while operating the frequency converter at the exact planned wavelength combination. This tuning method is incompatible with cryogenic operation since the temperature is fixed through the cryostat and the tunability in the low Kelvin range becomes negligible. As a consequence, we rely on accurate simulation and fabrication of the required cryogenic poling period.

\subsection{Cryogenic challenges in the Ti:PPLN material platform} \label{sec:cryochallenges}
Cryogenic operation of Ti:PPLN requires consideration of thermal effects which influence the nonlinear interaction. We see from Eq.~(\ref{eq:phasemismatch2}) that defining the poling period requires exact knowledge of the effective refractive indices and the thermal contraction of the material. The available refractive index data for bulk lithium niobate covers only operation temperatures from room temperature upwards~\cite{edwards1984temperature,jundt1997temperature}. We extrapolate this data for the cryogenic temperature range and feed it into our simulation. Empirical data for the thermal contraction is available for temperatures down to \SI{60}{K}~\cite{scott1989properties}. We assume the length remains constant for lower temperatures because the thermal expansion coefficient approaches zero at \SI{0}{K}. We see from our earlier studies on type-II phase-matching that this extrapolation approach introduces a deviation between simulation and experimental data for the low temperature range, which requires an empirical correction term~\cite{Lange2022cryogenic,lange2023degenerate}. Since the type-II process involves TE and TM polarization, we cannot distinguish between the impact of each polarization independently on the total deviation. In this paper, we use the same simulation data as before to investigate the uncertainties that apply if we only consider TM polarized light.

Beyond the need for accurate cryogenic refractive index data, photorefractive and pyroelectric effects bring further challenges for cryogenic operation. Photorefraction and pyroelectricity in lithium niobate can cause a localized accumulation of electric charges, triggered by different events~\cite{weis1985lithium}. Photorefraction is mainly driven by impurity centers such as $\mathrm{Fe}^{2+}$ and $\mathrm{Fe}^{3+}$ which are trapped during the growth process of the crystal~\cite{pal2015photorefractive}. These ions weakly absorb the in-coupled light which results in separation and redistribution of the charge carriers. In bright areas, the electrons are excited into the conduction band and then diffuse into the darker regions~\cite{pal2015photorefractive}. The resulting electric fields cause a variation in the refractive index profile of the waveguide due to the electro-optic effect. Especially for high power applications in the visible or near-infrared regime, this effect can cause beam distortion during propagation, such as defocusing, and thus reduce the overall device performance~\cite{rams2000optical}. Photorefraction was observed for TM- and TE-polarized modes, but the change in the extraordinary refractive index (affecting TM polarization) is much larger compared to the ordinary one (affecting TE polarization)~\cite{chen1969optically}. It was further shown that photorefractive damage can be characterized by an intensity threshold. This threshold increases for higher operation temperatures due to the increased electron mobility~\cite{rams2000optical,carrascosa2008understanding}. Consequently, we assume that photorefraction has a strong effect on cryogenic operation. The photo-excited electrons are likely to experience a ``freeze in'' and cannot move back into the waveguide region because of their low mobility.

Since lithium niobate is a pyroelectric material, it exhibits a spontaneous polarization that is oriented along the crystal axis. The strength of this polarization is temperature dependent; a change in temperature directly results in a change of the spontaneous polarization~\cite{lang2005pyroelectricity,jachalke2017measure}. For $z$-cut lithium niobate, this means that charge carriers accumulate on the top and bottom surface of the waveguide chip during the cooldown. Under ambient conditions, atmospheric charge carriers from the surroundings can neutralize these accumulated charges. The vacuum in a cryostat prevents this compensation, such that the accumulated charges can lead to localized electric fields and thus modifications of the refractive index. Previous work shows that during cooling to cryogenic temperatures, unpredictable changes can be measured in the electrical and optical properties of lithium niobate~\cite{thiele2020cryogenic,bartnick2021cryogenic,bravina2004low,thiele2024pyroelectric}. These bear the experimental signatures of spontaneous electric discharges of the unbound charge carriers, which take place on different time scales during the temperature transition. It was also observed that these electrical and optical perturbations often show a temporal correlation~\cite{thiele2024pyroelectric}. Most of these perturbations happen for temperatures above \SI{100}{K}. For lower temperatures the number of events reduces significantly which might be due to the reduced charge carrier mobility and a reduced pyroelectric coefficient~\cite{thiele2024pyroelectric}. Similar to photorefractive damage, any accumulated charges that do not discharge during the cooldown can be ``frozen in'' and will then disturb the cryogenic device performance.

As a consequence, at cryogenic temperatures we expect two main contributions that might alter our phase-matching properties. On the one hand the temperature change itself can result in localized refractive index modifications, and on the other hand, any in-coupled light intensity can cause additional perturbations when operating the frequency converter at a fixed cryogenic temperature. 

An additional issue that must be considered is the coupling to the waveguide in the cryostat. We take advantage of the low confinement of our waveguides; this allows us to employ a free-space coupled cryostat. This method introduces a large working distance of about \SI{13}{mm} between the lens and waveguide end facet. However, because of the large waveguide mode sizes, we do not rely on a particularly high numerical aperture which allows us to work with standard aspheric lenses. The cooldown process will cause a slight movement of the waveguide, mainly because the sample mount thermally contracts during temperature cycling. To compensate for this effect, the sample is mounted on piezo positioners to realign the waveguide. The free-space coupling thus represents a flexible method since the alignment can be readjusted at any time and we are able to reposition the sample to characterize several waveguides within one cooldown cycle.

\section{Experimental methods}
\begin{figure} [t]
    \centering
    \includegraphics[width=0.90\linewidth]{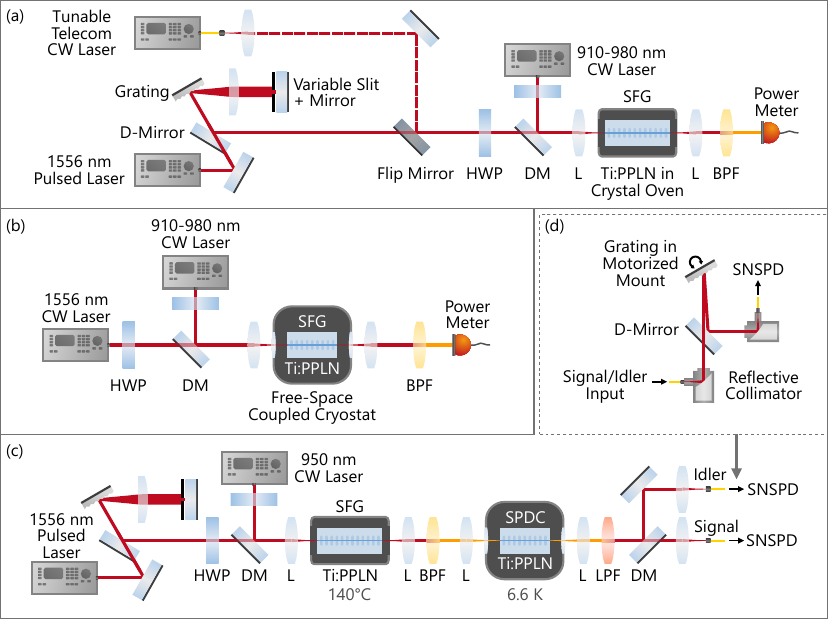}
    \caption{Schematic overview of the experimental setups to characterize the frequency conversion process under different conditions. We first perform sum frequency generation (SFG), (a) in a heated waveguide, and (b) in the cryogenic waveguide. (c) Next, we combine both setups to pump cryogenic spontaneous parametric down-conversion (SPDC) with the generated SFG. (d) The spectral properties of signal and idler are characterized using a grating spectrometer setup. D-Mirror: D-shaped mirror, HWP: half-wave plate, DM: dichroic mirror, L: aspheric lens, Ti:PPLN: periodically poled titanium in-diffused waveguides in lithium niobate, BPF: band-pass filter, LPF: long-pass filter, SNSPD: superconducting nanowire single-photon detector.}
    \label{fig:experimentalSetup}
\end{figure}

The experimental setup to characterize the cryogenic waveguide performance is shown in Fig.~\ref{fig:experimentalSetup}. We employ two in-house fabricated Ti:PPLN waveguide samples. The first waveguide is used to combine two lasers via SFG to generate the pump light for the following SPDC process. The pump light generation is depicted in Fig.~\ref{fig:experimentalSetup}~(a). The waveguide is positioned inside a crystal oven and heated to about \SI{140}{\degreeCelsius}. 
The chip has a total length of \SI{40}{mm} that is periodically poled with a period of \SI{9.11}{\um}. The end facets are anti-reflection (AR) coated, enabling high transmission for about \SI{950}{nm} and \SI{1550}{nm} on the in-coupling side and high transmission around \SI{590}{nm} on the out-coupling side. 

We couple two laser beams into the waveguide. The first laser is a continuous-wave (CW) laser that can be tuned from \SI{910}{nm} to \SI{980}{nm} with a maximum power of about \SI{80}{mW} (CTL 950, TOPTICA Photonics). In a first test, this laser beam is combined with a second tunable CW laser emitting in the telecom C-band (T500S, EXFO) for characterizing the phase-matching. Next, the second laser is exchanged by a pulsed femtosecond laser with a central wavelength of \SI{1556}{nm}, a spectral bandwidth of \SI{12}{nm}, and a repetition rate of \SI{80}{MHz} (Onefive Origami 15 LP, NKT Photonics). The pulsed laser is spectrally filtered with a folded $4f$-setup to decrease its spectral bandwidth while increasing the temporal bandwidth. This process increases the generated SFG intensity by better matching the interacting bandwidths and increasing the temporal overlap with the CW laser. The $4f$-setup consists of a blazed grating, a plano-convex lens, and a slit with variable width placed directly in front of a mirror. 
The polarization in each path is controlled with a half-wave plate, before both lasers are overlapped using a dichroic mirror. For the waveguide coupling we employ two aspheric lenses. 
We use two short-pass filters at \SI{850}{nm} to filter out the pump light. In addition, we set up a band-pass filter, consisting of one short-pass filter at \SI{650}{nm} and one long-pass filter at \SI{550}{nm} to ensure any higher harmonic light is filtered out as well. We detect the generated SFG light with a standard power meter.

The second waveguide is positioned inside a free-space coupled cryostat that provides optical access to the end facets through transparent windows. The base temperature at the sample stage is about \SI{6.6(2)}{K}. To characterize the phase-matching, we employ a similar setup as for the heated waveguide (see Fig.~\ref{fig:experimentalSetup}~(b)). The cryogenic chip has a length of \SI{22.9}{mm} that is poled with a period of \SI{9.64}{\um} and its end facets have the same AR coatings. We couple both lasers into the chip, this time using an achromatic, aspheric lens with long working distance (focal length of \SI{15}{mm}), as we are limited by the distance between the waveguide chip and the lens. 
The generated SFG light is out-coupled with another AR-coated aspheric lens. 
We use the same filter combination and power meter as before to detect the SFG.

In order to operate the cryogenic waveguide as an SPDC source, both setups are combined to pump this waveguide with the pulsed SFG signal (see Fig.~\ref{fig:experimentalSetup}~(c)). The photon pairs are filtered by long-pass filters at \SI{850}{nm} and separated by a dichroic mirror that cuts at \SI{1180}{nm}. Signal and idler are then coupled into polarization-maintaining (PM) single-mode fibers each and detected by superconducting nanowire single-photon detectors (SNSPDs) located in a separate cryostat. Both detectors and the attached fibers in the cryostat are optimized for an operation wavelength of about \SI{1550}{nm}. This limits the detection performance for the idler photons around \SI{950}{nm}. In order to characterize the spectral properties of the SPDC photons, we employ two home-built scanning-grating spectrometers as shown in Fig.~\ref{fig:experimentalSetup}~(d). The signal, or idler photons are coupled out of the fiber by a silver-coated reflective collimator. The diffractive optic is a blazed grating optimized for \SI{1600}{nm} and \SI{750}{nm} for the signal and idler photons respectively. The grating is used in Littrow configuration and the back-reflected wavelength components are coupled into a PM fiber by using a D-shaped mirror and a second collimator. We scan the angle of the grating to resolve an angle-dependent intensity spectrum. The calibration is performed by connecting a tunable CW laser and measuring transmitted intensity with a photo diode. The two spectrometers have unequal transmission bandwidths of $\Delta\lambda_\mathrm{s}=\SI{0.92(1)}{nm}$ and $\Delta\lambda_\mathrm{i}=\SI{0.52(3)}{nm}$ because their gratings have a different number of grooves. Further details on the spectrometer setup design can be found in the Supplemental Material of Ref.~\cite{Lange2022cryogenic}.

\section{Sum frequency generation for pump light generation}
As a first step, we characterize the properties of the heated waveguide used for SFG. This waveguide is fabricated the same way as the cryogenic waveguide under test, only differing in absolute length and poling period. Therefore, this initial characterization provides insights  into the nonlinear waveguide's behavior during active heating. These findings will serve as a point of comparison for the cryogenic waveguide performance.

We start to characterize the phase-matching spectra by performing CW SFG (compare Fig.~\ref{fig:experimentalSetup}~(a), flip mirror is inserted). This way, we can measure the spectra with higher resolution, as using the pulsed operation (flip mirror out), which will be used for the pump light generation. We set the wavelength of the CW telecom laser to \SI{1556.4}{nm} which corresponds to the center wavelength of the pulsed laser. The wavelength of the second laser is scanned and the generated SFG power is detected. We detect six peaks in the SFG intensity, corresponding to six phase-matched processes involving different spatial modes. For each mode, we set the laser to the peak wavelength, re-optimize the in-coupling, and perform a detailed scan. Furthermore, we take a mode image, using a 50x magnifying objective and focusing on a beam camera. 

\begin{figure} [t]
    \centering
    \includegraphics[width=1.0\linewidth]{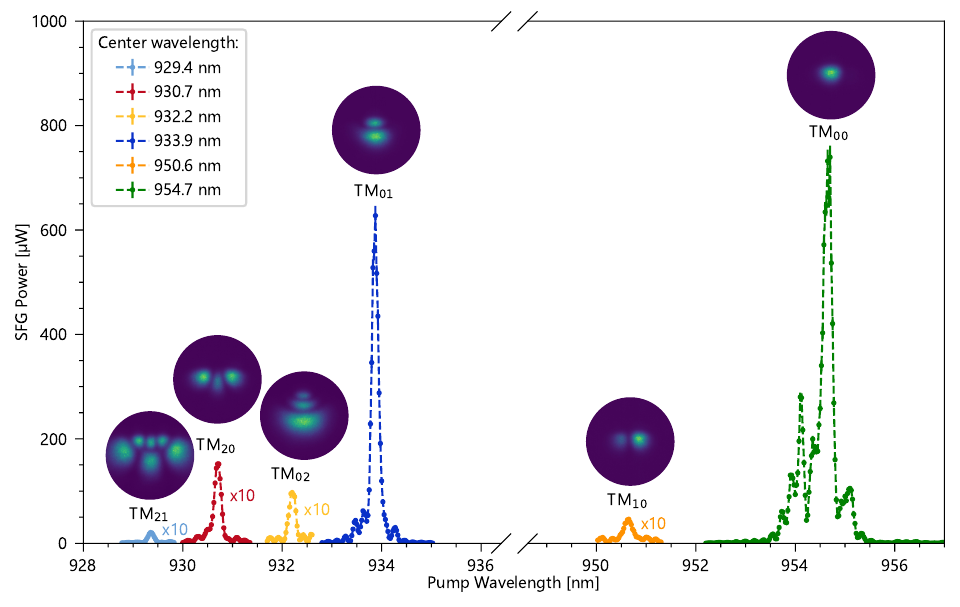}
    \caption{Phase-matched SFG processes in the heated waveguide operated at \SI{175}{\degreeCelsius}. We can clearly identify six phase-matched processes within the scanning range. The insets show images of the generated spatial modes. Note, that the intensity of the four smaller peaks was multiplied by a factor of 10 to make the data more clearly visible.}
    \label{fig:higherModes_Experiment}
\end{figure}
The measured phase-matching spectra and the spatial mode profiles are shown in Fig.~\ref{fig:higherModes_Experiment}. The spatial modes can be clearly identified from the taken images. This shows that the use of two CW lasers enable the generation of a specific spatial mode, which wavelength is given by conservation of energy. The slight variation in the phase-matching point, given by momentum conservation, is explained by slight differences in the effective refractive indices (see Fig.~\ref{fig:higherModes_RSoft}). We see maximum intensities for the $\mathrm{TM}_{01}$ and the $\mathrm{TM}_{00}$ modes. Due to the low intensity of the other modes, we neglect these interactions for the phase-matching considerations of the SPDC process. While we concentrate on the phase-matched process of all three fundamental modes, our results show that the generated $\mathrm{TM}_{01}$ mode reaches comparable intensity. This is because this spatial mode has a larger overall mode size than the fundamental mode at \SI{590}{nm}, which results in a large overlap of \SI{34.8}{\%} with the fundamental modes of the other beams. Due to the multi-mode behavior of the \SI{590}{nm} beam, it is challenging to couple only to the fundamental mode of our SPDC waveguide. This means that any intensity that will be coupled to the $\mathrm{TM}_{01}$ mode can result in SPDC photons at different wavelengths than our designed process. 

For the pump light generation, we pick the waveguide which shows highest SFG intensity in the fundamental mode. We exchange the CW telecom laser with the pulsed laser and tune the temperature of the SFG chip to \SI{140}{\degreeCelsius}, so that the phase-matching point is adjusted for generating \SI{590}{nm}. At this point we additionally tune the slit width in our $4f$-setup to maximize the generated SFG power. We obtain \SI{46.8}{\uW} when measuring an average input pump power of \SI{74.12}{mW} for the CW \SI{950}{nm} beam and \SI{2.31}{mW} for the pulsed \SI{1556.4}{nm} beam. We calculate the external nonlinear conversion efficiency by 
\begin{equation}
    \eta = \frac{P_\mathrm{out}}{P_\mathrm{in,1}P_\mathrm{in,2}L^2} \,,
\end{equation}
where $P_\mathrm{out}$ is the out-coupled SFG power, $P_\mathrm{in,1}$ and $P_\mathrm{in,2}$ are the in-coupled power values, and $L$ is the poled waveguide length. This results in a conversion efficiency of \SI{1.71}{\% / W cm^2}. This efficiency is limited by the definite spatial overlap of the differently sized fundamental waveguide modes and the temporal overlap of the pulsed with the CW laser. The generated SFG signal will be used to pump the cryogenic SPDC source. The beam has a center wavelength of \SI{590.016(2)}{nm} and a full width at half maximum (FWHM) of \SI{0.150(3)}{nm}, measured with a single-photon sensitive spectrometer. Despite limited efficiency, the heated waveguide enables stable SFG operation, indicating the absence of photorefractive damage at the employed optical power levels.

\section{Cryogenic waveguiding and phase-matching characterization}
We characterize the cryogenic waveguide with regard to the linear waveguiding quality and nonlinear phase-matching properties. We repeat the CW phase-matching investigation by performing SFG in the cryogenic waveguide, using the setup shown in Fig.~\ref{fig:experimentalSetup}~(b). We make several observations that will affect the SPDC process.

\subsection{Cryogenic waveguide performance}
We experience that the waveguides support guiding of the fundamental TE-polarized modes, while coupling of the TM-polarized modes is challenging. This applies to both CW lasers. The TE mode can be often clearly identified by observing the collimated mode profile on a fluorescent card. The TM mode however, is often not visible at all after rotating the half-wave plate to adjust the polarization. The centers of the mode distributions for the TE and TM modes are at slightly different heights, which explains why polarization rotation does not result in optimized coupling for the orthogonal polarization. However, at room temperature this is sufficient to achieve basic coupling, so that further alignment can be performed by optimizing on this transmitted mode. At cryogenic temperatures, we sometimes rely on inspecting the transmitted intensity with the 50x magnifying objective and a beam camera, before we can perform further alignment. Even after careful optimization, the intensity which is transmitted through the waveguide for TM-polarized light, remains very small. 

We quantify the transmission ${T=P_\mathrm{out}/P_\mathrm{in}}$ by detecting the input power $P_\mathrm{in}$ in front of the in-coupling lens and the transmitted power $P_\mathrm{out}$ behind the out-coupling lens. The measured transmission values are summarized in Table~\ref{tab:waveguidetransmission}. We measure the room temperature transmission for TM-polarized light and show the values for the cryogenic transmission for each polarization. The cryogenically detected transmission for TM polarization is much lower than at room temperature. The cryogenic transmission for TE polarization is higher than for the TM-polarized beam. 
\begin{table} [t]
    \centering
    \caption{Measured waveguide transmission values $T$ for the two lasers at \SI{950}{nm} and \SI{1556}{nm}. The subscripts denote the operation temperature (RT - room temperature, CT - cryogenic temperature) and the polarization (TM, TE).}
    \label{tab:waveguidetransmission}
    \begin{tabular}{ccccc}
        \hline
        Wavelength [nm]& $T_{\mathrm{RT,TM}}$ [\%] & $T_{\mathrm{CT,TM}}$ [\%] & $T_{\mathrm{CT,TE}}$ [\%]\\ \hline
        950 & 43.4 & 10.6 & 21.5 \\
        1556 & 53.6 & 24.9 & 37.6 \\
        \hline
    \end{tabular}
\end{table}
\begin{figure} [t]
    \centering
    \includegraphics[width=1\linewidth]{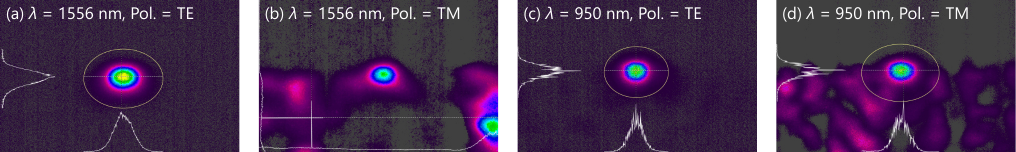}
    \caption{Spatial mode images of the cryogenic waveguide for the two CW lasers and both polarizations. The intensity of the guided TM modes is in the same order of magnitude as surrounding substrate modes, while the TE modes are clearly separated from the background. 
    Note that the images were directly saved by the beam camera software (BeamMic, Ophir), thus including an overlay, marking the area that was recognized as the mode position. Raw data was saved within this overlay region only.}
    \label{fig:modeImages}
\end{figure}

However, the measured values for the transmitted power include not only light intensity that is guided through the waveguide, but also any substrate modes which still reach the detector. This is important to note since we observe that the cryogenic TM-polarized waveguide mode is accompanied by significant substrate modes, while the TE-polarized waveguide mode is separated much stronger from potential scattered background. This can be seen from the mode images shown in Fig.~\ref{fig:modeImages} which were measured together with the cryogenic transmission values. 
We assume that this stray light is mainly caused by pyroelectric charges that arise during the cooldown. These charges move along the crystalline $c$-axis and thus induce a change in the spatial distribution of the extraordinary refractive index which is parallel to this axis. However, the ordinary refractive index, which is experienced by the TE-polarized light, might be largely unaffected as this polarization is aligned perpendicular to the arising electric fields. This is consistent with observations of photorefractive damage, where changes in the extraordinary refractive index are known to be significantly larger than those in the ordinary index~\cite{chen1969optically,pal2015photorefractive}. 

We see from Fig.~\ref{fig:modeImages} that the brightness of the TM-polarized waveguide modes is in the same order of magnitude as stray light in the surrounding substrate. 
Consequently, we observe a significant drop in detected TM-polarized power when placing a pinhole in front of the power meter, which transmits primarily the collimated waveguide mode. The TE-polarized power remains largely unaffected by inserting the pinhole. We thus conclude that the transmission measurements for TE-polarized light express the actual waveguide power more accurately than for TM polarization. The absolute TM-polarized power in the waveguide mode is much lower as it seems by inspecting the overall transmission. For this reason, the detected transmission values together with the mode images verify that the cryogenic operation has a strong impairing effect on TM polarized light.

\subsection{Variations in the cryogenic phase-matching}
\begin{figure} [t]
    \centering
    \includegraphics[width=0.86\linewidth]{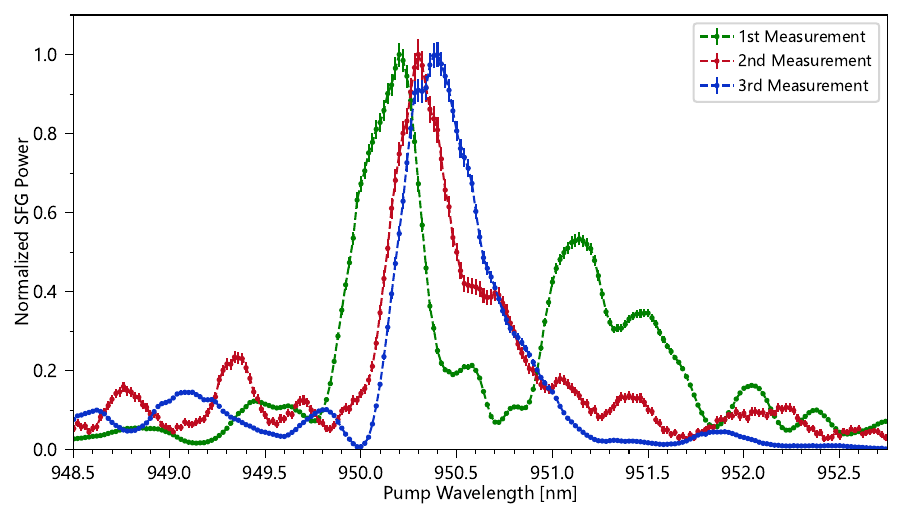}
    \caption{Variations in the phase-matching of the cryogenic waveguide. The power of the SFG signal was normalized to better represent the changes in the spectral shape. Measurements two and three were taken about a month after measurement one, during a separate cooldown.}
    \label{fig:phaseMatchingChanges}
\end{figure}
In addition to the inconsistencies of the waveguiding performance, we observe fluctuations in the cryogenic phase-matching. We characterize the phase-matching spectrum of our waveguide three times. The changes in the spectrum are shown in Fig.~\ref{fig:phaseMatchingChanges}. The first measurement was performed in a separate cooldown than the other two measurements. Measurement two was carried out about one month later in a subsequent cooldown. Measurement three was performed about two days later, but within the same cooldown cycle. In all three measurements, the spectral shape shows deviations from the ideal $\mathrm{sinc}^2$-shape since the main peak is accompanied by irregular side peaks of varying intensity. Compared to the heated SFG spectra, the cryogenic waveguide consistently experiences more significant phase-matching distortions across all measurements. The maximum cryogenic conversion efficiency in these measurements was \SI{0.88}{\% / W cm^2} when generating \SI{2.4(2)}{\uW} SFG power, which is about a factor of two smaller than the value measured at \SI{140}{\degreeCelsius}. We note that we cannot compare the two values since they were measured with different waveguide samples and the cryogenic value was measured for CW operation while the heated value was measured when overlapping the CW and pulsed laser. Moreover, we cannot rule out differences in the coupling efficiency. However, even though we are not limited by the temporal overlap of both lasers when pumping with CW only, the cryogenic value is lower than the pulsed heated efficiency. This indicates that the overall efficiency is reduced under cryogenic conditions. We also see that the general spectral shape varies between all three measurements and the main peak shows a slight wavelength shift. Since the overall shift is only about \SI{0.3}{nm}, this still shows that the central wavelength is reproducible within a small window through our measurements.

\subsection{Consequences of the cryogenic operation}
The classical characterization of our cryogenic waveguide, including measurements of transmission, the mode profile, and phase-matching of the SFG process, clearly reveals limitations of the guiding quality for TM polarization under cryogenic conditions. While the mode images for TE polarization show a clear mode profile, the TM mode was often accompanied by a lot of stray light in the substrate. The extent of this effect varied between different cooldown cycles. This points to an unpredictable impact of pyroelectric charges which are ``frozen in'' during the cooldown and do not discharge once the sample is operated at cryogenic temperatures. These charges are likely to remain in the same place as long as they are not excited by adding external energy, either in the form of thermal heat, or by optical exposure. We observed this behavior in a waveguide kept under cryogenic conditions for a week without exposure to light, which showed almost no changes in the phase-matching spectrum when tested with low optical input power.

As we have also observed variations in the phase-matching spectra within one cooldown cycle, the cooldown implications cannot be explained solely by pyroelectricity. The performance changes appearing while the waveguide was kept at a constant temperature can be attributed to photorefractive damage. This was verified by detecting the peak power of the cryogenic SFG signal with increasing input power. First, the SFG power increases with pump power, but at a certain threshold the efficiency drops again. When only coupling the higher energy pump laser at \SI{950}{nm} to the waveguide, the transmitted power increased linearly with the in-coupled power. This points to the conclusion that for the tested input power range (up to about \SI{45}{mW}), the pump lasers are not the limiting factors considering photorefraction. The damage is probably induced by the generated \SI{590}{nm} beam which has much higher photon energy. Accordingly, the cryogenic threshold for photorefractive damage in the visible range appears to be in the low microwatt range, as we observed phase-matching changes when generating only a few microwatt of SFG power. This clearly limits the input power that can be coupled to the waveguide in the following SPDC process. Moreover, because we observed no change in the phase-matching shape over one week, we can infer that charge accumulation is not reversible within this time at cryogenic temperatures. This suggests that induced damage can build up over repeated measurements.

\section{Cryogenic spontaneous parametric down-conversion}
Following the cryogenic SFG measurements, we combine the heated SFG pump light generation and the cryogenic waveguide to perform the SPDC experiment (see Fig.~\ref{fig:experimentalSetup}~(c)). The pulsed SFG beam at \SI{590}{nm} is coupled through the cryogenic waveguide and the generated signal and idler photons are detected with SNSPDs. We refer to the photons in the telecom range as the signal photons and to the photons centered around \SI{950}{nm} as the idler photons. 

\subsection{Pump-power dependent performance}
\begin{figure} [t] 
    \centering
    \includegraphics[width=1\linewidth]{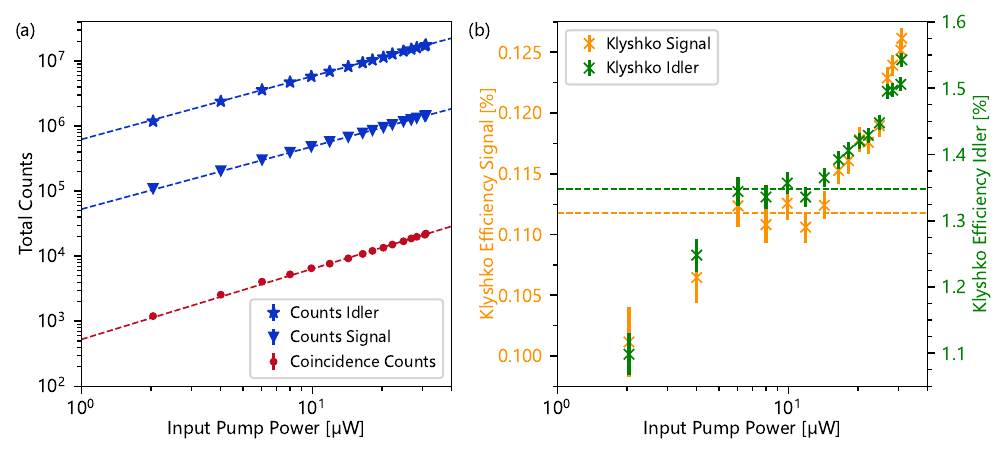}
    \caption{Measured number of counts from the cryogenic SPDC source for increasing pump power. (a) Total counts of signal, idler, and coincidences for an integration time of \SI{60}{s}. (b) Klyshko efficiency of signal and idler, calculated from the measured counts. The dashed lines serve as a guide to the eye.}
    \label{fig:klyshkoEfficiency}
\end{figure}

We measure the total clicks of the signal and idler detector and the coincidence counts between both. This is done in dependence of the pump power for an integration time of \SI{60}{s}. This data is shown in Fig.~\ref{fig:klyshkoEfficiency}~(a). We use a logarithmic scale on the y-axis to display all count rates on the same axis. From this, it can be directly seen that there is a significant difference in the signal and idler count rate. We detect about a factor of 10 more idler counts around \SI{950}{nm} than signal counts at about \SI{1556}{nm}. This is unexpected since both SNSPDs are optimized for the signal photon's wavelength in the telecom range. Consequently, we would expect the signal path to have lower losses than the idler path. The difference in counts is thus more likely to be caused by additional noise photons reaching the idler detector, than significant loss in the signal path.

We can support this consideration by the following observation. The number of counts in the idler arm can be increased even more when changing the polarization of the pump beam from TM to TE polarization. This results in a strong increase of idler counts, while we see a significant drop of the signal and coincidence counts. The polarization dependence points to another nonlinear frequency conversion process which is phase-matched when pumping the cryogenic waveguide with TE-polarized light. We assume that this process yields one photon which reaches the idler detector. The wavelength of the signal photon however, might be too long to either be properly coupled to the optical fiber which is connected to the SNSPD, or to trigger a detection event in case it reaches the detector. Since this process yields a high idler count rate, these photons can artificially increase the detected idler counts when the polarization is set to TM. Slight imperfections, for example in the employed half-wave plate, can cause a remnant of TE polarization which might excite this process. 

The nonlinear coefficient for lithium niobate is the highest for a TM-polarized type-0 process. Therefore, a process pumped with TE polarization has a lower nonlinear coefficient, according to the material parameters~\cite{sharapova2017toolbox}. However, as we have seen from our linear cryogenic characterization, the waveguiding itself is strongly degraded for TM-polarized light. Consequently, cryogenic operation could affect the TM-polarized photons so significantly that TE-pumped interactions achieve higher experimental efficiency. For this reason, thorough investigation of a cryogenic TE-polarized type-0 interaction should be part of future work beyond the scope of this paper.

Apart from the difference in measured counts, we see in Fig.~\ref{fig:klyshkoEfficiency}~(a) that all counts increase almost linearly for increasing input power. This shows that the available pulsed input power of about \SI{30}{\uW} does not induce additional photorefractive damage as this would cause a saturation, or even reduction of the counts for high pump power. This could indicate that the cryogenic waveguide demonstrates improved tolerance to a coupled beam in the visible range when operated in pulsed rather than CW mode. 

From our measurement, we calculate the Klyshko efficiency that represents the losses in the signal and idler arm. The efficiency of the signal arm is given by the coincidences $C_\mathrm{s,i}$ divided by the idler counts $C_\mathrm{i}$, according to
\begin{equation}
    \eta_\mathrm{Klyshko,s} = \frac{C_\mathrm{s,i}}{C_\mathrm{i}} \,.
\end{equation}
The input power dependent Klyshko efficiencies for both arms are shown in Fig.~\ref{fig:klyshkoEfficiency}~(b). The overall efficiencies are very low. The calculated Klyshko efficiency for the signal arm is especially low because it is artificially reduced by the noise counts in the idler arm. In the small pump power regime around \SI{10}{\uW} we can recognize a slight plateau before the values increase for increasing pump power. We refer this power regime to the contribution of first-order photon pairs while the mean photon number increases beyond this range for stronger pump power. The first two measurement points are below the plateau. We assume this to be a sign of a constant noise background that adds up to the power dependent nonlinear noise counts. In the low pump power regime, this constant noise background will be dominant. Consequently, the reduced signal-to-noise ratio results in a decreased Klyshko value at low pump power. The overall small Klyshko efficiencies indicate that a large amount of photons already gets lost inside the waveguide due to the reduced waveguiding. This is also evident in a reduced effective length, which we derive from the following spectral measurement.

\subsection{Joint spectral intensity}
\begin{figure} [t]
    \centering
    \includegraphics[width=0.90\linewidth]{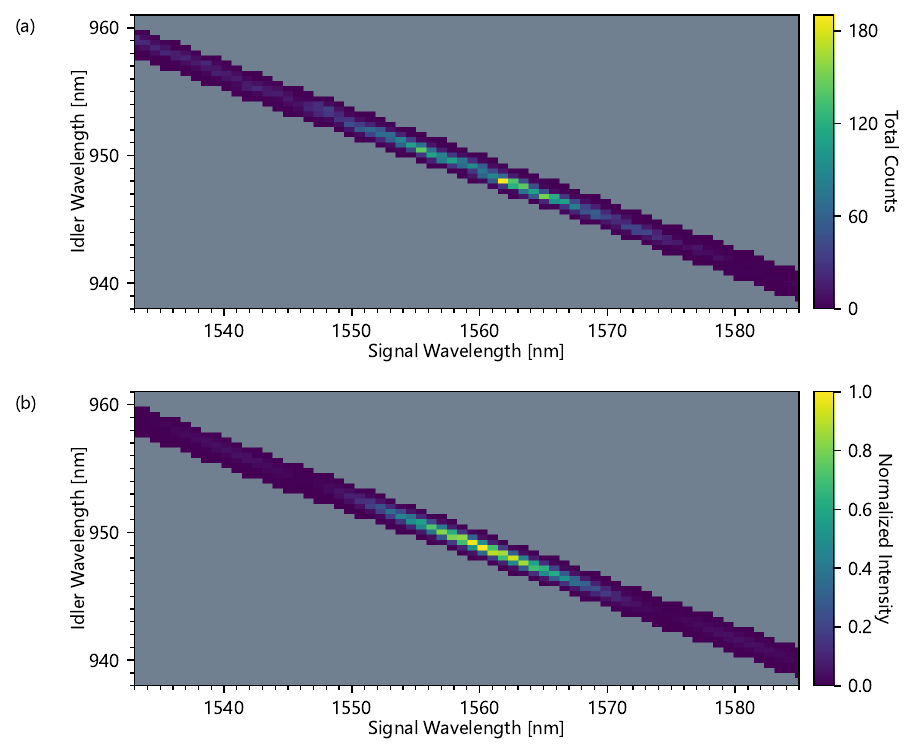}
    \caption{Joint spectral intensity of the cryogenic SPDC source. (a) Measured total coincidence counts, summarized for two measurements with an integration time of \SI{80}{s} each. The measurement range was adjusted to cover a linear section around the pump distribution function. No data was recorded for the gray area. (b) Simulation of the JSI for an effective length of \SI{3.8(3)}{mm} and a poling period of \SI{9.654(2)}{\um}.}
    \label{fig:JSI_Experiment}
\end{figure}
In addition to the signal and idler counts, we investigate their correlated spectral properties by measuring the joint spectral intensity (JSI). This is done by inserting two spectrometer setups as shown in Fig.~\ref{fig:experimentalSetup}~(d) into the signal and idler path, respectively. We scan the angles of the two gratings step by step while detecting the number of coincidence counts. The JSI must be limited to the area of the pump distribution, which is given by energy conservation. We thus restrict our measurement array to a small linear region covering the expected JSI distribution. Since we have to detect every pixel individually, this reduces the measurement time significantly. We perform the same measurement twice, integrating for \SI{80}{s} at every measurement point. Both measurements were performed within the same cooldown cycle with a break of one day in between. The results were very similar, indicating no significant changes due to photorefraction during the measurement time. We sum up the total counts of both measurements and show the resulting JSI in Fig.~\ref{fig:JSI_Experiment}~(a). 

We simulate the expected cryogenic JSI using the effective refractive index data as discussed in Section~\ref{sec:cryochallenges}. The simulation further includes the spectral profile of the pump beam, the fabricated poling period, the resolution of the spectrometers, and the effective length over which all three fields are interacting. We perform the simulation according to the method which is described in the Supplemental Material of Ref.~\cite{Lange2022cryogenic}. To account for the spectrometer resolution, we first perform this simulation with higher resolution than the measurement. This simulated JSI is then convolved with the Gaussian transmission profile of each spectrometer along the two axes, respectively. Finally, the JSI is down-sampled to display the result with the measurement's resolution. In the first simulation, we keep all parameters fixed to the design. This yields a JSI for which the center wavelengths are slightly shifted in comparison to the measurement result. We present the central wavelengths in Table~\ref{tab:centerwavelengths}. The offset can be explained either by a slight variation between the extrapolated refractive indices and the experimental ones, or by small fabrication uncertainties in the waveguide dimensions or poling period. Since the total shift is only within a few nanometers, this shows that the extrapolated refractive indices describe the process already very well and no significant correction is required for TM polarization. 
\begin{table} [t]
    \centering
    \caption{Overview of the center wavelengths of the signal and idler photons. The values are extracted from a Gaussian fit that was applied to the projection of the JSI on the x- and y-axis, respectively. The experimental values are taken from the JSI shown in Fig.~\ref{fig:JSI_Experiment}~(a). Simulation~1 is performed using the design parameters. Simulation~2 considers optimization of the effective length and the poling period to fit the experiment, as shown in Fig.~\ref{fig:JSI_Experiment}~(b).}    
    \begin{tabular}{cccc}
        \hline
         & Experiment & Simulation 1 & Simulation 2  \\ \hline
        Signal Wavelength [nm] & $1560.4 \pm 0.3$ & $1552.31 \pm 0.04$ & $1560.44 \pm 0.04$ \\
        Idler Wavelength [nm] & $948.6 \pm 0.2$ & $951.78 \pm 0.02$ & $948.75 \pm 0.02$ \\ \hline
    \end{tabular}

    \label{tab:centerwavelengths}
\end{table}

We perform a second simulation with an optimization algorithm. We allow variations in the effective length and the poling period and thus minimize the deviation of the simulated JSI and our measured distribution. This way, we can estimate the effective length of our waveguide which affects the broadening of the JSI along the axis of the pump distribution function. We allow a slight deviation of the poling period, to let both JSIs overlap at the same central wavelengths to get a more accurate result for the effective length. We show this simulation in Fig.~\ref{fig:JSI_Experiment}~(b) and the center wavelengths in Table~\ref{tab:centerwavelengths}. According to this optimization, the effective length of our waveguide is \SI{3.8(3)}{mm}, which corresponds to about \SI{16.6}{\%} of the designed length which was periodically poled during fabrication. We expect this reduction to be caused by a significant amount of pyroelectric and photorefractive charges, which have accumulated during the previous experiments. The optimized poling period is \SI{9.654(2)}{\um}, which deviates by only \SI{0.15}{\%} from the designed period of \SI{9.64}{\um}. This deviation is thus within fabrication tolerances. While the reduced effective length reflects the limited cryogenic waveguide performance, the minimal deviation of the poling period verifies the accuracy of our theoretical model and that we can reliably achieve phase-matching for the designed wavelength combination.

\section{Conclusion}
We demonstrated a thorough investigation of the cryogenic performance of titanium in-diffused waveguides in lithium niobate. We characterized a type-0 nonlinear interaction that employs TM-polarized fields only, to take advantage of the highest nonlinear coefficient. Although this process is expected to show highest conversion efficiency, this polarization experiences stronger degradation due to photorefractive and pyroelectric damage than TE-polarized light. By characterizing the cryogenic SFG spectrum and the SPDC performance we obtained deeper insights into the cryogenic material properties and possible performance boundaries. 

Our results show that cryogenic operation strongly affects the TM-polarized light and causes slight, unpredictable variations in the linear mode guiding and nonlinear efficiencies. While the waveguides remained efficient for TE-polarized light, the transmitted intensity for TM polarization was strongly reduced. Moreover, we saw that a significant amount of the TM-polarized intensity was coupled to substrate modes. This reduced waveguiding quality resulted in a short effective length and thus low nonlinear conversion efficiencies. We observed variations in the absolute performance metrics between different cooldowns, but also within the same cooldown cycle. These oscillations were caused by both the spontaneous accumulation of electric charges during the cooldown procedure, and by a low threshold for the coupled power in the visible range. Furthermore, we saw that pumping the SPDC source with visible light is challenging due to the multi-mode nature of the \SI{590}{nm} beam. Light intensity that is coupled to a spatial mode apart from the fundamental will not contribute to the designed SPDC process. We further experienced a significant amount of noise counts in the SPDC process, which we assume to be caused by additional phase-matched processes.

Despite the drawbacks that limit the cryogenic performance, the central wavelengths of the SFG and SPDC process were reproducible within a small window. Even though there were variations in the performance metrics, the overall phase-matching point can be achieved with high accuracy. The experimental poling period that we derived from our measurements differs by only \SI{0.15}{\%} from our design, which verifies an accurate theoretical model and precise waveguide fabrication.

Our characterization revealed that titanium in-diffused waveguides in lithium niobate can in principle be employed for cryogenic nonlinear frequency conversion, however, limited to small optical power values, low conversion efficiency, and with minor but unpredictable variations in the performance metrics. 
Future investigations could involve a cryogenic TE-polarized type-0 interaction, pulsed cryogenic operation only, and enhanced spectral filtering of SPDC photons to prevent the detection of photons from additional phase-matching. 
Furthermore, a thorough comparison of the cryogenic performance to room temperature operation could be of interest. However, for a fixed poling period, the phase-matched wavelengths shift significantly during the cooldown. According to our simulation, the center wavelengths for signal and idler at room temperature for the characterized waveguide are approximately \SI{1700}{nm} and \SI{900}{nm}. Therefore, a like-for-like comparison at room temperature of the same waveguide and the same combination of pump laser and SFG frequency is not directly possible.

In conclusion, we anticipate that significant engineering is required to overcome the aforementioned challenges for efficient cryogenic nonlinear frequency conversion in this material platform. Nevertheless, research in other platforms, such as LNOI, is already addressing these issues. In this context, our results offer an initial benchmark against which future cryogenic integration platforms can be compared.

\newpage
\begin{backmatter}
\bmsection{Funding}
We acknowledge financial support from the Deutsche Forschungsgemeinschaft (Grant No. 231447078–TRR 142).

\bmsection{Disclosures}
The authors declare no conflicts of interest.

\bmsection{Data Availability Statement}
Data underlying the results presented in this paper are available in Ref.~\cite{lange_2025_16911614}.

\end{backmatter}

\bibliography{library}

\end{document}